\documentclass[12pt]{article}
\usepackage{epsfig}
\newcommand{\tdm}[1]{\mbox{\boldmath $#1$}}

\newcommand{\eq}{\begin{equation}}
\newcommand{\eqx}{\end{equation}}
\newcommand{\eqn}{\begin{eqnarray}}
\newcommand{\eqnx}{\end{eqnarray}}
\newcommand{\dt}{\Delta}

\begin{document}
%
\begin{center}
{\large\bf QCD expectations for the spin structure function $g_1$ in the
low $x$ region}
\vspace{1.1cm}\\
{\sc J.~Kwieci\'nski,}\footnote{e-mail: jkwiecin@solaris.ifj.edu.pl}
{\sc B. Ziaja}\footnote{e-mail: beataz@solaris.ifj.edu.pl}\\
\vspace{0.3cm}
{\it Department of Theoretical Physics, \\
H.~Niewodnicza\'nski Institute of Nuclear Physics,
Cracow, Poland}
\end{center}
\vspace{1.1cm}
\begin{abstract}
The structure function $g_1(x,Q^2)$ is analysed within the
formalism based on unintegrated spin dependent parton
distributions  incorporating the LO Altarelli-Parisi evolution and
the double $ln^2(1/x)$ resummation at low $x$. We
quantitatively examine possible role of the latter for the small
$x$ behaviour of the structure functions $g_1$ of the nucleon
within the region which may be relevant for the possible
polarized HERA measurements. We show that while the non-singlet
structure function is dominated at low $x$ by ladder diagrams
the contribution of the non-ladder bremsstrahlung terms is
important for the  singlet structure function.  Predictions for
the polarized gluon distribution $\Delta G(x,Q^2)$ at low $x$
are also given.
\end{abstract}
\vspace{1.1cm}
\section{Introduction}
Understanding of the small $x$ behaviour of the spin dependent structure functions
of the nucleon, where $x$ is the Bjorken parameter is interesting
both theoretically and phenomenologically.
Present experimental measurements  do not cover the very low values of $x$ 
and so
the knowledge
of  reliable extrapolation of the structure functions
into this region is important for  estimate of  integrals which appear
in the Bjorken and Ellis-Jaffe sum rules \cite{ALTAR}.  Theoretical description
of the structure function $g_1^p(x,Q^2)$ at low $x$ is also extremely
relevant for the possible polarised HERA measurements \cite{ALBERT}.

The purpose of the present paper is to explore the theoretical
QCD expectations concerning the small $x$ behaviour of the spin
dependent structure functions taking into account the double
$ln^2(1/x)$ resummation.  The dominant contribution generating
these
double logarithmic terms is given by the ladder diagrams with
the quark (antiquark) and gluon exchanges along the ladder.  The very
transparent way of resumming these terms is provided by the
formalism of the unintegrated (spin dependent) parton
distributions which satisfy the corresponding integral
equations.  In this paper we extend this formalism so as to
include the non-ladder bremsstrahlung terms.  We argue that
these terms can be easily included by adding the suitable higher order
contributions to the kernels of the corresponding integral equations.
We also incorporate within this scheme the complete LO
Altarelli-Parisi evolution thus obtaining the unified system of
equations
which makes it possible to analyse simultaneously the parton
distributions in the large and small $x$ regions.  
In particular, this formalism allows us to extrapolate dynamically
the spin dependent structure functions from the region of large and
moderately small values of  $x$,  where they are constrained by
the presently available data to the (very) small $x$ domain
which can possibly be probed at the polarized HERA.

The content of our paper is as follows.
In the next section we summarize the expectations of the Regge
pole model for the small $x$ behaviour of the spin dependent
structure functions.  We point out in particular potentially
small magnitude ($\sim 0$) of the corresponding Regge pole
intercepts which control the small $x$ behaviour of the spin
dependent structure functions. We shall also remind  that the
Regge pole model expectations with the corresponding intercepts
($\sim 0$) become unstable against the
conventional QCD evolution which generates more singular
behaviour at small $x$. In Section 3 we present the discussion
of the double $ln^2(1/x)$ terms using the formalism of the
unintegrated parton distributions.  These distributions will be
the basic quantities which will satisfy the corresponding
integral equations generating the double $ln^2(1/x)$ resummation.
We shall formulate these equations for the sum of ladder
diagrams and extend the formalism by including the non-ladder
bremsstrahlung terms.  This will be done by a suitable
modification of the kernel(s) of the corresponding integral
equations.  For fixed QCD coupling the equations will generate
the complete small $x$ behaviour of spin dependent structure
functions.  The formalism will be further extended by allowing the
coupling constant to run and by including the complete LO
Altarelli Parisi evolution.  We shall discuss the analytic
structure
of the solution due to these modifications.
In Sec. 4 we will present numerical solution of the integral
equations
starting from the simple semi-phenomenological
parametrization of the non-perturbative part of the spin
dependent parton distributions. Finally in Sec. 5 we give
summary of our results.

\section{Regge pole model expectations for the small $x$ behaviour
of $g_1$ and LO perturbative QCD effects}

The small $x$ behaviour of spin dependent structure functions
for fixed $Q^2$ reflects the high energy behaviour of
the
virtual Compton scattering (spin dependent) total cross-section with
increasing total CM energy
squared $W^2$ since $W^2=Q^2(1/x-1)$. This is, by definition, the Regge limit
and so the Regge pole exchange picture
\cite{PC} is
therefore quite appropriate for the theoretical description
of this behaviour. Here as usual $Q^2=-q^2$ where $q$ is the four momentum transfer
between
the scattered leptons. The relevant Reggeons which describe the small
$x$ behaviour of the spin dependent structure functions
are those which correspond to the axial vector mesons \cite{IOFFE,KARL}.

The Regge pole model gives the following small $x$ behaviour of the structure
functions $g_1^{i}(x,Q^2)$~:
\begin{equation}
g_1^{i}(x,Q^2) = \gamma_i(Q^2)x^{-\alpha_{i}(0)},
\label{rg1}
\end{equation}
where $g_1^{i}(x,Q^2)$
denote either  singlet  ($g_1^s(x,Q^2)=
g_1^p(x,Q^2)+g_1^n(x,Q^2)$)  or non-singlet ($g_1^{ns}(x,Q^2)=
g_1^p(x,Q^2)-g_1^n(x,Q^2)$) combination of structure functions.

\noindent
In Eq.\ (\ref{rg1})
$\alpha_{s,ns}(0)$ denote the intercept of the Regge pole
trajectory corresponding to the axial vector mesons with $I=0$
or $I=1$ respectively. It is expected that   $\alpha_{s,ns}(0) \le
0$ and that $\alpha_{s}(0) \approx \alpha_{ns}(0)$ i.e. the singlet
spin dependent structure function is expected to have similar low $x$
behaviour as the non-singlet one in the Regge pole model.

Several of the Regge pole model expectations
for spin dependent structure functions
are modified by  perturbative QCD effects.  In particular the Regge behaviour
(\ref{rg1}) with $\alpha_i(0)\leq 0$ becomes unstable against the QCD evolution which generates
more singular behaviour than that given by Eq. (\ref{rg1}) for $\alpha_{i}(0)\le 0$.  In the LO approximation one gets:
$$
g_1^{NS}(x,Q^2) \sim exp [2 \sqrt{\Delta P_{qq}(0) \xi(Q^2) ln(1/x)}],
$$
\begin{equation}
g_1^S(x,Q^2) \sim  exp [2 \sqrt{\lambda_0^{+} \xi(Q^2) ln(1/x)}],
\label{dlogq2}
\end{equation}
where
\begin{equation}
 \xi(Q^2) = \int_{Q_0^2}^{Q^2}{dq^2\over q^2} {\alpha_s(q^2)\over 2 \pi}
\label{xi}
\end{equation}
and
\begin{equation}
\lambda_0^+ = {\Delta P_{qq}(0) +\Delta P_{gg}(0) +
\sqrt{(\Delta P_{qq}(0) -\Delta P_{gg}(0))^2 +
4\Delta P_{qg}(0)\Delta P_{gq}(0)}\over 2}
\label{lplus}
\end{equation}
with $\Delta P_{ij}(0) = \Delta P_{ij}(z=0)$, where $\Delta P_{ij}(z)$
denote the LO splitting functions describing evolution
of spin dependent  parton densities. To be precise we have~:
\begin{eqnarray}
{\bf \dt P(0)} \equiv
\left( \begin{array}{cc} \Delta P_{qq}(0) & \Delta P_{qg}(0)\\
\Delta P_{gq}(0) & \Delta P_{gg}(0) \\ \end{array} \right ) =
\left( \begin{array}{cc}  {{N^2-1 \over 2N}} & - N_F \\
                          {{N^2-1 \over  N}} &   4N  \\ \end{array} \right),
\label{dpij}
\end{eqnarray}
where $N$ denotes number of colours, and $N_F$ denotes number of flavours.
\section{Double logarithmic $ln^2(1/x)$ corrections to $g_1$}

The LO (and NLO) QCD evolution which sums the leading (and next-to-leading)
powers of $ln(Q^2/Q_0^2)$ is incomplete at low $x$.  In this region
one should worry about another "large" logarithm which is $ln(1/x)$ and
resum its leading powers. In the spin independent case this is provided
by the Balitzkij, Fadin, Kuraev, Lipatov (BFKL) equation \cite{BFKL} which
gives in the leading $ln(1/x)$ approximation the following small $x$ behaviour
of the structure function $F_1^S(x,Q^2)$~:
\begin{equation}
F_1^S(x,Q^2) \sim x^{-\lambda_{BFKL}},
\label{bfkl1}
\end{equation}
where the intercept of the BFKL pomeron $\lambda_{BFKL}$ is given in the leading order by the
following formula:
\begin{equation}
\lambda_{BFKL}=1+{3\alpha_s\over \pi}4ln(2).
\label{bfkl2}
\end{equation}
It has recently been pointed out that the spin dependent structure function
$g_1$ at low $x$ is dominated  by the {\it double} logarithmic $ln^2(1/x)$
contributions
i.e. by those terms of the perturbative expansion which correspond to
the powers of $ln^2(1/x)$ at each order of the expansion
\cite{BARTNS,BARTS}.  Those contributions go {\it beyond}
the LO or NLO order QCD evolution of polarized parton densities \cite{AP} and in order to take
them into account one has to
include the resummed double $ln^2(1/x)$ terms in the coefficient
and splitting functions \cite{RG,JAP}.
In the following we will  discuss an alternative approach to the double
$ln^2(1/x)$ resummation based on unintegrated distributions
\cite{BBJK,BZIAJA,MAN}.

To this aim we introduce the unintegrated (spin dependent) parton
distributions
$f_i(x^{\prime},k^2)$ ($i=u_{v},d_{v},\bar u,\bar d,\bar s,g$) where $k^2$
is the transverse momentum squared of the parton $i$ and $x^{\prime}$ the
longitudinal momentum fraction of the parent nucleon carried by a parton.
Those functions will satisfy the corresponding linear integral equations
generating the double $ln^2(1/x)$ resummation.

The conventional (integrated) distributions $\Delta p_i(x,Q^2)$ are
related in the
following way  to the unintegrated distributions $f_i(x^{\prime},k^2)$:
\begin{equation}
 \Delta p_i(x,Q^2)=\Delta p_i^{(0)}(x)+
 \int_{k_0^2}^{W^2}{dk^2\over k^2}f_i(x^{\prime}=
x(1+{k^2\over Q^2}),k^2),
\label{dpi}
\end{equation}
where $\Delta p_i^{(0)}(x)$ is the nonperturbative part 
of the distribution.
The parameter $k_0^2$ is the infrared cut-off which will be set equal
to 1 GeV$^2$.
The nonperturbative part $\Delta p_i^{(0)}(x)$ can be viewed
upon as originating from the integral over non-perturbative region  $k^2<
k_0^2$, i.e.
\begin{equation}
\Delta p_i^{(0)}(x)= \int_{0}^{k_0^2}{dk^2\over k^2}f_i(x,k^2).
\label{gint0}
\end{equation}
The spin dependent structure function $g_1^p(x,Q^2)$ of the proton is
related in a standard way to the (integrated) parton distributions:
$$
g_1^p(x,Q^2)=
$$
\begin{equation}
{1\over 2}\left[{4\over 9}(\Delta u_v(x,Q^2) + 2\Delta \bar u (x,Q^2))+
{1\over 9}(\Delta d_v(x,Q^2) + 2\Delta \bar u(x,Q^2) +  2\Delta \bar s(x,Q^2))
\right],
\label{gp1}
\end{equation}
where $\Delta u_v(x,Q^2) \equiv \Delta p_{u_v}(x,Q^2)$ etc.
In what follows we assume that $\Delta \bar u =\Delta \bar d$
and set the number of flavours $N_F$ to be equal to three.

It is convenient to consider separately the valence quark
distributions and the assymetric part of the sea~:
\begin{equation}
f_{us}(x^{\prime},k^2)= f_{\bar u}(x^{\prime},k^2)-f_{\bar s}(x^{\prime},k^2)
\label{us}
\end{equation}
which do not couple with the gluons, and the singlet
distribution~:
\begin{equation}
f_{\Sigma}(x^{\prime},k^2) = f_{u_v}(x^{\prime},k^2)+f_{d_v}(x^{\prime},k^2)+
4f_{\bar u}(x^{\prime},k^2)+
2f_{\bar s}(x^{\prime},k^2)
\label{sns}
\end{equation}
together with the gluon distribution $f_g(x^{\prime},k^2)$
which satisfy the coupled integral equations.

The full contribution to the double $ln^2(1/x)$ resummation comes from
the ladder diagrams with quark and gluon exchanges along the ladder
(see Figs.\ 1,2) and the non-ladder bremsstrahlung diagrams (Fig.\ 3). The latter ones are
obtained from the ladder diagrams by adding to them soft bremsstrahlungs gluons
or soft quarks \cite{BARTNS,BARTS} and they generate the infrared corrections
to the ladder contribution.
\begin{figure}[t]
    \centerline{
     \epsfig{figure=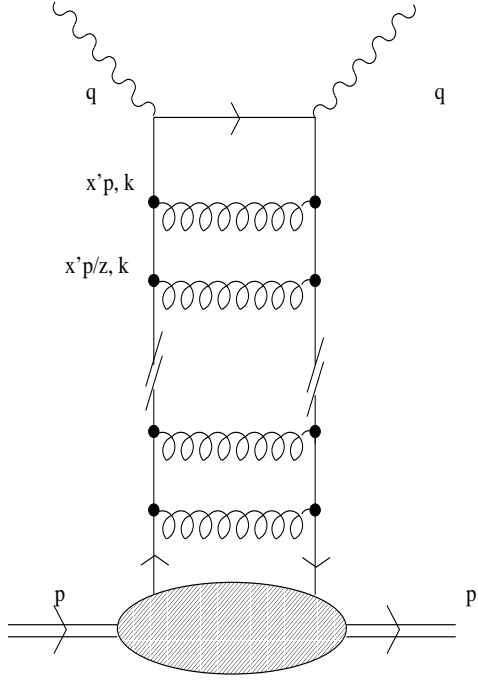,height=9cm,width=6cm}
               }
     \caption{Ladder diagram generating
     double logarithmic terms in the  non-singlet spin structure function
     $g_1^{NS}$.}
\label{fig.1}
\end{figure}
%
\begin{figure}[t]
    \centerline{
     \epsfig{figure=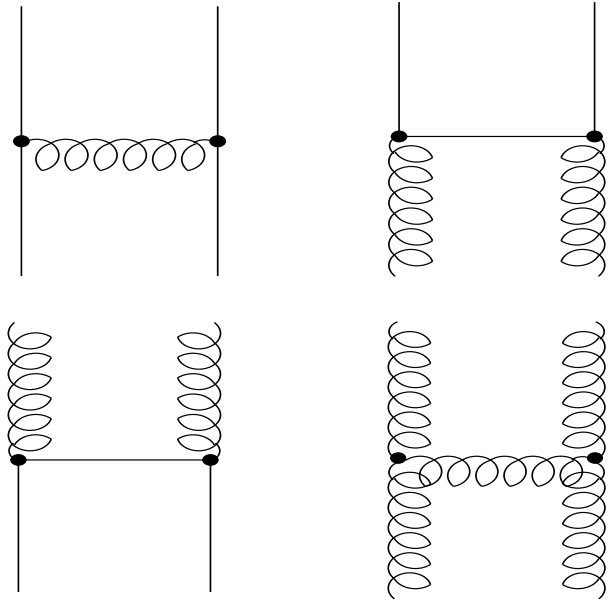,height=8cm,width=8cm}
               }
     \caption{Ladder elements for a singlet case}
\label{fig.2}
\end{figure}
%
\noindent
\begin{figure}[t]
    \centerline{
     \epsfig{figure=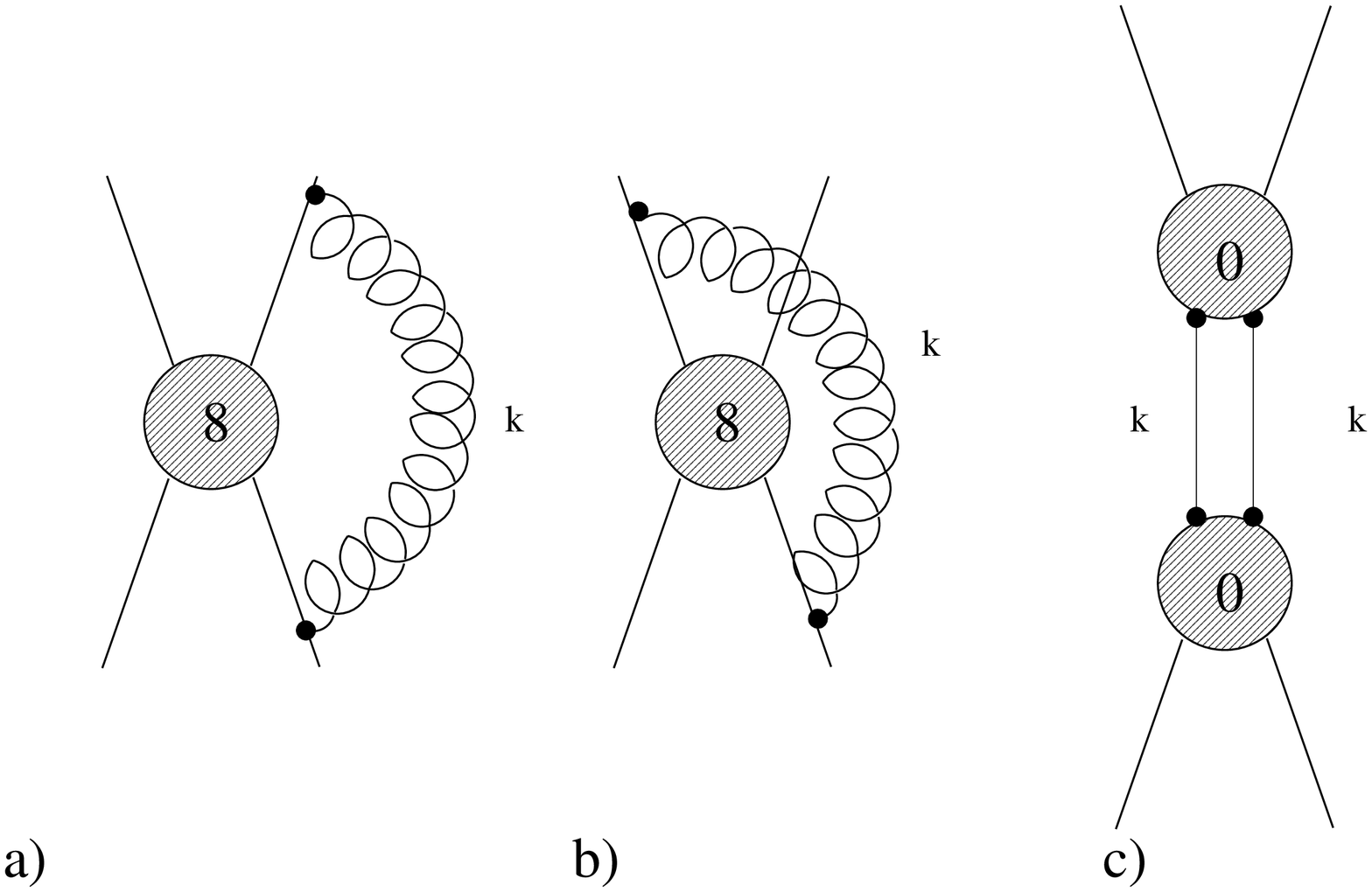,height=8cm,width=15cm}
               }
\caption{Non-ladder contributions containing a), b) soft gluon or c) soft quark
attached to the quark-quark scattering amplitude}
\label{fig3}
\end{figure}
%
%
%
The sum of double logarithmic $ln^2(1/x)$
terms corresponding to ladder diagrams is generated by the following
integral equations (see Figs.1, 2)~:
\begin{equation}
f_{j}(x^{\prime},k^2)=f^{(0)}_{j}(x^{\prime},k^2) +
{ \alpha_s \over 2 \pi} \Delta P_{qq}(0)
\int_{x^{\prime}}^1 {dz\over z}
\int_{k_0^2}^{k^2/z}
{dk^{\prime 2}\over k^{\prime 2}}
f_{j}\left({x^{\prime}\over z},k^{\prime 2}\right),
\label{dlxns}
\end{equation}
where $j= u_v, d_v, us$, and~:
{\footnotesize
\eqn
f_{\Sigma}(x^{\prime},k^2)=f^{(0)}_{\Sigma}(x^{\prime},k^2)+
{\alpha_s\over 2 \pi}
\int_{x^{\prime}}^1 {dz\over z}
\int_{k_0^2}^{k^2/z}
{dk^{\prime 2}\over k^{\prime 2}}
\left[\Delta P_{qq}(0)
f_{\Sigma}\left({x^{\prime}\over z},k^{\prime 2}\right)+
\Delta P_{qg}(0)
f_{g}\left({x^{\prime}\over z},k^{\prime 2}\right)\right]\nonumber,
\eqnx}
{\footnotesize
\eqn
f_{g}(x^{\prime},k^2)=f^{(0)}_{g}(x^{\prime},k^2) +
{\alpha_s\over 2 \pi}
\int_{x^{\prime}}^1 {dz\over z}
\int_{k_0^2}^{k^2/z}
{dk^{\prime 2}\over k^{\prime 2}}
\left[\Delta P_{gq}(0)
f_{\Sigma}\left({x^{\prime}\over z},k^{\prime 2}\right)+
\Delta P_{gg}(0)
f_{g}\left({x^{\prime}\over z},k^{\prime 2}\right)\right]\nonumber\\
\label{dlxsg}
\eqnx}
with $\Delta P_{ij}(0) \equiv \Delta P_{ij}(z=0)$ given by Eq.\ (\ref{dpij}).

The variables $k^2$($k^{\prime 2}$) denote the transverse momenta
squared of the
quarks (gluons) exchanged along the ladder, $k_0^2$ is  the infrared cut-off
and the inhomogeneous terms
$f_i^{(0)}(x^{\prime},k^2)$ will be specified later.
The integration limit $k^2/z$
follows from the requirement that the  virtuality of the
quarks (gluons) exchanged along the ladder is controlled by
the tranverse momentum squared.

Equation (\ref{dlxns}) is similar to the corresponding equation in QED
describing the double logarithmic resummation generated by ladder diagrams
with fermion exchange \cite{QED}. The problem of double logarithimc
asymptotics in QCD in the non-singlet channels was also discussed in 
Ref.\ \cite{QCD,NSQCD,JK}.

Equations (\ref{dlxns}, \ref{dlxsg}) generate singular power behaviour of the
spin dependent parton distributions and of the spin dependent
structure functions $g_1$  at small  $x$ i.e.
$$
g_1^{NS}(x,Q^2) \sim x^{-\lambda_{NS}},
$$
$$
g_1^{S}(x,Q^2) \sim x^ {-\lambda_{S}},
$$
\begin{equation}
\Delta G(x,Q^2) \sim x^ {-\lambda_{S}},
\label{power}
\end{equation}
where $g_1^{NS}=g_1^p-g_1^n$ and $g_1^{S}=g_1^p+g_1^n$ respectively, and
$\Delta G$ is the spin dependent gluon distribution.  The behaviour 
of spin structure functions reflects
the small $x^{\prime}$ behaviour of their unintegrated distributions.
Exponents $\lambda_{NS,S}$ are given by the following formulas:
\eq
\lambda_{NS} = 2 \sqrt{\left[{\alpha_s \over 2\pi} \Delta P_{qq}(0)\right]},
\label{lambdans}
\eqx
\begin{equation}
\lambda_{S} = 2 \sqrt{\left[{\alpha_s \over 2\pi} \lambda_0^+\right]},
\label{lambda}
\end{equation}
where $\lambda_0^+$ is given by Eq.\ (\ref{lplus}). The power-like behaviour
(\ref{power}) with the exponents $\lambda_{NS,S}$ given
by Eq.\ (\ref{lambda}) remains
the leading small $x$ behaviour of the structure functions provided that
their non-perturbative parts are less singular.  This takes place
if the latter are assumed to have the Regge pole like behaviour with the
corresponding intercept(s) being near $0$.
%
\begin{figure}[t]
    \centerline{
     \epsfig{figure=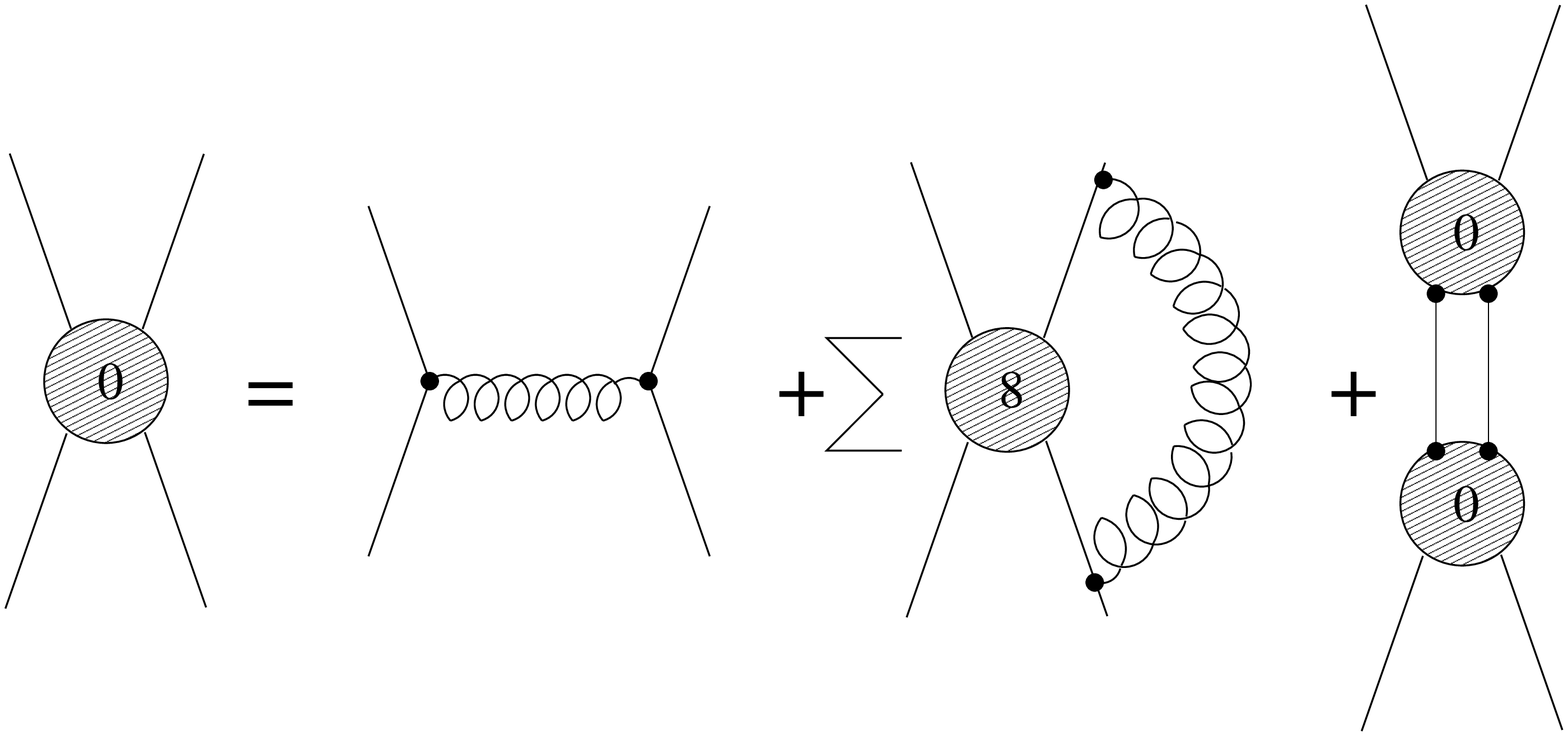,height=7cm,width=14cm}
               }
     \caption{Infrared evolution equation for partial wave matrix $F_0$}
\label{fig.3}
\end{figure}
The exponents $\lambda_{NS}$ and $\lambda_S$ correspond to the
leading singularities of the moment functions $\bar f_i(\omega,k^2)$
( $\Delta \bar p_i(\omega,
Q^2)$ ) for $i={\small NS} (i=u_v, d_v, us)$ and $i=\Sigma,i=g$ respectively in
the $\omega $ plane.  The moment functions $\bar f_i(\omega,
k^2)$  and $\Delta \bar p_i(\omega,
Q^2)$ are defined in a conventional way~:
\\
$$
\bar f_i(\omega,k^2) = \int_0^1 dx^{\prime} x^{\prime \omega - 1}
f_i(x^{\prime},k^2),
$$
\begin{equation}
\Delta \bar p_i(\omega,Q^2)=\int_0^1 dx x^{ \omega - 1} \Delta
p_i(x,Q^2).
\label{moments}
\end{equation}
Ladder equations (\ref{dlxns}), (\ref{dlxsg}) in $\omega-$space take then
the form~:
{\footnotesize
\eqn
\bar f_i(\omega,k^2)&=&\bar f_{i}^{(0)}(\omega,k^2)+
{\bar \alpha_s \over \omega}
\dt P_{qq}(0)
\left[\int_{k_0^2}^{k^2}
{dk^{\prime 2}\over
k^{\prime 2}}\bar f_i(\omega,k^{\prime 2})+
\int_{k^2}^{\infty} {dk^{\prime 2}\over
k^{\prime 2}}\left({k^2 \over k^{\prime 2}}\right)^{\omega}
\bar f_i(\omega,k^{\prime 2})\right],
\label{dleqm}
\\
{\bf \bar f}_S(\omega,k^2)&=&{\bf \bar f}_{S}^{(0)}(\omega,k^2)+
{\bar \alpha_s \over \omega}
{\bf M}_0
\left[\int_{k_0^2}^{k^2}
{dk^{\prime 2}\over
k^{\prime 2}}{\bf \bar f}_S(\omega,k^{\prime 2})+
\int_{k^2}^{\infty} {dk^{\prime 2}\over
k^{\prime 2}}\left({k^2 \over k^{\prime 2}}\right)^{\omega}
{\bf \bar f}_S(\omega,k^{\prime 2})\right],
\label{dleqms}
\eqnx
}
where $i=u_v,d_v,us$, and $\bar \alpha_s$ is defined as~:
\eq
\bar \alpha_s=\frac{\alpha_s}{2\pi},
\eqx
and~:
\eqn
{\bf \bar f}_{S}(\omega,k^2)&=& \left( \begin{array}{cc} \bar f_{\Sigma}(\omega,k^2)\\
\bar f_g(\omega,k^2) \\ \end{array} \right ),\nonumber\\
{\bf M}_0 \equiv {\bf \dt P}(0)&=&\left( \begin{array}{cc} \Delta P_{qq}(0) & \Delta P_{qg}(0)\\
\Delta P_{gq}(0) & \Delta P_{gg}(0) \\ \end{array} \right ).
\label{m0}
\eqnx

The inhomogeneous terms $\bar f_i^{(0)}$ and ${\bf
\bar f}_S^{(0)}$ are related to the moment functions 
$\Delta \bar  p_i^{(0)}(\omega)$
 of the nonperturbative parts $
\Delta p_i^{(0)}(x)$ and ${\bf
\Delta p}_S^{(0)}(x)$ of the integrated distributions
in the following way~:
\eqn
\bar f_i^{(0)}(\omega,k^2)&=&
\bar \alpha_s {\Delta P_{qq}(0)\over \omega} \Delta \bar  p_i^{(0)}(\omega),
\nonumber\\
{\bf \bar f}_S^{(0)} (\omega,k^2) &=&
\bar \alpha_s {{\bf M_0}\over \omega} {\bf \Delta \bar  p}_S^{(0)}(\omega).
\label{inho1}
\eqnx
For fixed coupling $\alpha_s$ one may obtain analytic solution of (\ref{dleqm}),
(\ref{dleqms})~:
\begin{equation}
\bar f_i(\omega,k^2)=R_i(\bar \alpha_s, \omega)
\left({k^2\over k_0^2}\right)^{\gamma_{NS}(\alpha_s,\omega)},
\label{solm}
\end{equation}
\begin{equation}
{\bf \bar f}_S(\omega,k^2)=
\left({k^2\over k_0^2}\right)^{\tdm \gamma_{S}(\alpha_s,\omega)}
{\bf R}_S(\bar \alpha_s, \omega),
\label{solms}
\end{equation}
where $\gamma_{NS}$, ${\tdm\gamma_{S}}$ represent anomalous dimensions for
the non-singlet and singlet case respectively, and are equal to~:
\eqn
\gamma_{NS}(\omega,\alpha_s ) &=& {\omega - \sqrt{\omega^2 - 4 \dt P_{qq}(0)\bar
\alpha_s}\over 2}
\label{anomd}\\
{\tdm\gamma}_{S}(\omega,\alpha_s) &=& {\omega - \sqrt{\omega^2 - 4 {\bf M_0}
\bar \alpha_s}\over 2},
\label{anomds}
\eqnx
and  $R_i(\bar \alpha_s,\omega)$ and ${\bf R}_S(\bar\alpha_s,\omega)$
are given by the following equations~:
$$
R_i(\bar \alpha_s,\omega) = \gamma_{NS} (\bar \alpha_s,\omega)
\Delta \bar p_i^{(0)} (\omega)
$$
\begin{equation}
{\bf  R}_S(\bar \alpha_s,\omega) = {\tdm\gamma_{S}} (\bar \alpha_s,\omega)
{\bf \Delta \bar p}_S^{(0)} (\omega).
\label{fact1}
\end{equation}

It should be reminded that for the singlet case only the eigenvalues
of $\tdm\gamma_S$ contribute to the final solution. The solution takes
then the explicit form~:
\begin{equation}
\bar f_{\Sigma,g}(\omega,k^2) = R_{\Sigma,g}^{+}(\omega,\alpha_s,)
\left({k^2\over k_0^2}\right)^{\gamma^{+}(\omega,\alpha_s)}
                           + R_{\Sigma,g}^{-}(\omega,\alpha_s)
\left({k^2\over k_0^2}\right)^{\gamma^{-}(\omega,\alpha_s)},
\label{solsing}
\end{equation}
where $\lambda_0^+$, $\lambda_0^-$ denote eigenvalues of of $\tdm \gamma_S$~:
\begin{equation}
\gamma^{\pm}(\omega,\alpha_s)= {{\omega-\sqrt{\omega^2-4{\bar \alpha_s}
\lambda^{\pm}_0}}\over 2},
\label{gammapm}
\end{equation}
and $\lambda^+_0$ (the eigenvalue of ${\bf M}_0$) is defined
by equation (\ref{lplus}) while $\lambda^-_0$ reads:
\begin{equation}
\lambda^-_0 = {\Delta P_{qq}(0) +\Delta P_{gg}(0) -
\sqrt{(\Delta P_{qq}(0) -\Delta P_{gg}(0))^2 +
4\Delta P_{qg}(0)\Delta P_{gq}(0)}\over 2}.
\label{lminus}
\end{equation}
It can be shown that the moment functions $\Delta
\bar p_i(\omega,Q^2)$  and ${\bf \Delta \bar p}_S(\omega,Q^2)$ of the spin
dependent distributions have  the familiar RG form (for the fixed coupling)~:
$$
\Delta \bar p_i(\omega,Q^2)= \bar R_i(\bar \alpha_s,\omega) \left(
{Q^2\over k_0^2}\right)^{\gamma_{NS} (\bar \alpha_s,\omega)} + O(k_0^2/Q^2)
$$
\begin{equation}
{\bf \Delta \bar  p}_S(\omega,Q^2)=\left(
{Q^2\over k_0^2}\right)^{{\tdm\gamma_{S}} (\bar \alpha_s,\omega)}
{\bf \bar R}_S(\bar \alpha_s,\omega) + O(k_0^2/Q^2),
\label{rg}
\end{equation}
where
$$
\bar R_i(\bar \alpha_s,\omega)={\Gamma(\gamma_{NS} (\bar \alpha_s,\omega)+1)
\Gamma(\omega-\gamma_{NS} (\bar \alpha_s,\omega))\over
\Gamma(\omega)}
\Delta \bar  p_i^{(0)}(\omega)
$$
\begin{equation}
{\bf \bar R}_S(\bar \alpha_s,\omega)={\Gamma({\tdm\gamma_{S}}
(\bar \alpha_s,\omega)+1)
\Gamma(\omega-{\tdm\gamma_{S}} (\bar \alpha_s,\omega))\over
\Gamma(\omega)}
{\bf \Delta \bar  p}_S^{(0)}(\omega).
\label{prefact}
\end{equation}
To derive (\ref{rg}) we notice that equation (\ref{dpi}) 
implies the following relation between the moment functions 
$\Delta \bar p_i(\omega,Q^2)$ and $\bar f_i(\omega,k^2)$: 
\begin{equation}
\Delta \bar p_i(\omega,Q^2) = \Delta \bar p_i^{(0)}(\omega) + 
\int_{k_0^2}^{\infty} {dk^2\over k^2} \left(1+{k^2\over Q^2}\right)^{-\omega}
\bar f_i(\omega,k^2).
\label{int1}
\end{equation}
Substituting expressions (\ref{solm}) and (\ref{solms}) 
into (\ref{int1}) and taking into account equation (\ref{fact1}) we obtain
equation (\ref{rg}).       

Equations (\ref{anomd}), (\ref{anomds}) 
imply that the anomalous dimensions $\gamma_{NS}(\omega,\alpha_s)$
and $\gamma^{\pm}(\omega,\alpha_s)$ have the branch point
singularities in $\omega$ plane located at
$\omega=2\sqrt{\bar \alpha_s\Delta P_{qq}(0)}$ and
at $\omega=2\sqrt{\bar \alpha_s\lambda_0^{\pm}}$ respectively.
The small $x$ behaviour of the parton distributions and of the
structure functions  which is given by equations (\ref{power})
with the exponents $\lambda_{NS}$
and $\lambda_S$ defined by equations (\ref{lambdans}) and (\ref{lambda}) 
just reflects
the fact that this behaviour is controlled by the leading singularities in
the $\omega$ plane.

It may be seen from equations (\ref{rg},\ref{prefact}) that
the branch point singularities of anomalous dimensions  are also present in the
prefactors $\bar R_i(\bar \alpha_s,\omega)$ and ${\bf \bar
R}_S(\bar \alpha_s,\omega)$ and so the low $x$ behaviour of spin
dependent structure functions given
by equations (\ref{power}) is expected to hold for arbitrary values
of the scale $Q^2$.

Complete sum of double logarithmic $ln^2(1/x)$ terms  does also include
the non-ladder bremsstrahlung contributions besides the
ladder ones.  Expressions (\ref{lambdans}), (\ref{lambda}) corresponding
to the ladder diagrams are therefore only approximate.
The method of implementing the non-ladder bremsstrahlung corrections 
into the double logarithmic resummation was originally proposed by Kirschner, Lipatov \cite{QCD}, and applied
for the case of DIS small $x$ asymptotics in \cite{BARTNS,BARTS}, \cite{RG}.
Below we show how to implement these terms within our formalism.
In order to include the non-ladder corrections to the double logarithmic asymptotics
we use the infrared evolution equations derived in \cite{BARTNS,BARTS,QCD}.
The evolution equations for singlet partial waves $F_0$, $F_8$ (see Fig.\ 4 )~:
\eq
{\bf F}_{0,8}=\left( \begin{array}{cc} F_{qq}^{(0,8)} & F_{qg}^{(0,8)}\\
F_{gq}^{(0,8)} & F_{gg}^{(0,8)} \\ \end{array} \right )
\eqx
have  the form~:
\eqn
{\bf F}_0(\omega, \alpha_s)&=&\frac{8\pi^2\bar\alpha_s}{\omega}{\bf M}_0
             -\frac{4 \bar\alpha_s}{\omega^2}{\bf F}_8(\omega, \alpha_s){\bf G}_0
             +\frac{1        }{8\pi^2\omega}{\bf F}_0^2(\omega, \alpha_s)
\label{f0}\\
{\bf F}_8(\omega, \alpha_s)&=&\frac{8\pi^2\bar\alpha_s}{\omega}{\bf M}_8
             +\frac{\bar\alpha_s N}{\omega}\frac{d}{d\omega}{\bf F}_8(\omega,\alpha_s)
             +\frac{1        }{8\pi^2\omega}{\bf F}_8^2(\omega, \alpha_s),
\label{f8}
\eqnx
where ${\bf M}_0$ (cf.\ \ref{m0}), ${\bf M}_8$ are splitting functions matrices
in colour singlet and octet $t-$channel, and ${\bf G}_0$
contains colour factors resulting from attaching the soft gluon to external
legs of the scattering amplitude~:
\eqn
{\bf M}_8 &=&\left( \begin{array}{cc} -{1 \over 2N} & -{N_F \over 2}\\
                                                N & 2N \\ \end{array} \right ),
\nonumber \\
{\bf G}_0 &=&\left( \begin{array}{cc}  {N^2-1 \over 2N} & 0  \\
                                                    0 & N  \\ \end{array} \right ).
\label{m8g0}
\eqnx

Non-singlet contribution comes from $F_0^{qq}$ partial wave, as expected.
Only the partial waves ${\bf F}_0$ are contributing to the anomalous dimensions.
An explicit relation between partial waves and anomalous dimension reads~:
\eqn
{\bf F}_0   &=&8\pi^2 {\tdm\gamma}_S^{RES}\\
F_{qq}^{(0)}&=&8\pi^2 \gamma_{NS}^{RES}
\eqnx
and it depends already on resummed anomalous dimensions $\gamma^{RES}$
with all infrared non-ladder corrections included. Equation (\ref{f0})
for ${\bf F}_0$  may be solved analytically generating the following
expressions for anomalous dimensions~:
\eqn
\gamma_{NS}^{RES}(\alpha_s, \omega)&=& \frac{\omega}{2}
\Biggl(1-\sqrt{1-       \frac{4 \bar\alpha_s }{\omega}
\Biggl(                 \frac{\dt P_{qq}(0)}{\omega} -
   \frac{( {\bf F}_8(\omega){\bf G}_0)_{qq} }{2\pi^2\omega^2} \Biggr)}\Biggr)
\label{anomdres}\\
{\tdm\gamma}_{S}^{RES}(\alpha_s, \omega)&=& \frac{\omega}{2}
\Biggl(1-\sqrt{1-       \frac{4 \bar\alpha_s }{\omega}
\Biggl(                 \frac{{\bf M}_0}{\omega}
      -\frac{{\bf F}_8(\omega){\bf G}_0}{2\pi^2\omega^2} \Biggr)}\Biggr).
\label{anomdress}
\eqnx
Comparing the expressions for anomalous dimensions with and without 
the non-ladder contribution  
(\ref{anomd}), (\ref{anomds}), (\ref{anomdres}), (\ref{anomdress})
one may notice that the bremsstrahlung contribution
adds an additional term under the square root of $\gamma$'s~:
\eq
\sqrt{1- \frac{4 \bar\alpha_s  }{\omega        }
         \frac{ {\bf M}_0      }{\omega        }     }
\,\,\,\,\,{\bf \longrightarrow}\,\,\,\,\,
\sqrt{1- \frac{4 \bar\alpha_s  }{\omega        }
\Biggl(        \frac{ {\bf M}_0      }{\omega        }
    -\frac{ {\bf F}_8(\omega){\bf G}_0 }{2\pi^2\omega^2} \Biggr)   }.
\label{change}
\eqx
The same anomalous dimensions would be obtained if we modified
the kernels of  equations (\ref{dleqm}) and (\ref{dleqms}) by
setting $\bar \alpha_s \left ({{\bf M_0}\over \omega}-{{\bf
F_8}(\omega) {\bf G_0}\over 2 \pi^2 \omega^2}\right)$
in place of  $\bar \alpha_s {{\bf M_0}\over \omega}$.
The modified equations then read:
{\footnotesize
\eqn
\bar f_i(\omega,k^2)&=&\bar f_{i}^{(0)}(\omega,k^2)\nonumber\\
&+&\bar \alpha_s \left ({{\dt P(0)}\over \omega}-{{\bf 
F_8}(\omega) {\bf G_0}\over 2 \pi^2 \omega^2}\right)_{qq}
\left[\int_{k_0^2}^{k^2}
{dk^{\prime 2}\over
k^{\prime 2}}\bar f_i(\omega,k^{\prime 2})+
\int_{k^2}^{\infty} {dk^{\prime 2}\over
k^{\prime 2}}\left({k^2 \over k^{\prime 2}}\right)^{\omega}
\bar f_i(\omega,k^{\prime 2})\right],
\label{dleqmpopr}
\\
{\bf \bar f}_S(\omega,k^2)&=&{\bf \bar f}_{S}^{(0)}(\omega,k^2)\nonumber\\
&+&\bar \alpha_s \left ({{\bf M_0}\over \omega}-{{\bf 
F_8}(\omega) {\bf G_0}\over 2 \pi^2 \omega^2}\right)
\left[\int_{k_0^2}^{k^2}
{dk^{\prime 2}\over
k^{\prime 2}}{\bf \bar f}_S(\omega,k^{\prime 2})+
\int_{k^2}^{\infty} {dk^{\prime 2}\over
k^{\prime 2}}\left({k^2 \over k^{\prime 2}}\right)^{\omega}
{\bf \bar f}_S(\omega,k^{\prime 2})\right],
\label{dleqmspopr}
\eqnx
}
The corresponding equations for the functions
$f_i(x^{\prime},k^2)$ and for for ${\bf f}_S$ have the form~:
{\small
\eqn
f_{j}(x^{\prime},k^2)=f^{(0)}_{j}(x^{\prime},k^2) &+&
{\bar \alpha_s}\, \Delta P_{qq}(0)
\int_{x^{\prime}}^1 {dz\over z}
\int_{k_0^2}^{k^2/z}
{dk^{\prime 2}\over k^{\prime 2}}
f_{j}\left({x^{\prime}\over z},k^{\prime 2}\right)\nonumber\\
&-&{\bar \alpha_s}
\int_{x^{\prime}}^1 {dz\over z}
\Biggl(
\Biggl[ \frac{\tilde  {\bf F}_8 }{\omega^2} \Biggr](z)
\frac{ {\bf G}_0 }{2\pi^2}
\Biggr)_{qq}
\int_{k_0^2}^{k^2}
{dk^{\prime 2}\over k^{\prime 2}}
f_{j}\left({x^{\prime}\over z},k^{\prime 2}\right)\nonumber\\
&-&{\bar \alpha_s}
\int_{x^{\prime}}^1 {dz\over z}
\int_{k^2}^{k^2/z}
{dk^{\prime 2}\over k^{\prime 2}}
\Biggl(
\Biggl[\frac{\tilde   {\bf F}_8 }{\omega^2} \Biggr]
\Biggl(\frac{k^{\prime 2}}{k^2}z \Biggr)\frac{ {\bf G}_0 }{2\pi^2}
\Biggr)_{qq}
f_{j}\left({x^{\prime}\over z},k^{\prime 2}\right),
\label{dlres}
\eqnx
}
where $j= u_v, d_v, us$, and~:
{\small \eqn
{\bf f}_{S}(x^{\prime},k^2)={\bf f}^{(0)}_{S}(x^{\prime},k^2) &+&
{\bar \alpha_s}\, {\bf M}_0
\int_{x^{\prime}}^1 {dz\over z}
\int_{k_0^2}^{k^2/z}
{dk^{\prime 2}\over k^{\prime 2}}
{\bf f}_{S}\left({x^{\prime}\over z},k^{\prime 2}\right)\nonumber\\
&-&{\bar \alpha_s}
\int_{x^{\prime}}^1 {dz\over z}
\Biggl[ \frac{\tilde  {\bf F}_8 }{\omega^2} \Biggr](z)
\frac{ {\bf G}_0 }{2\pi^2}
\int_{k_0^2}^{k^2}
{dk^{\prime 2}\over k^{\prime 2}}
{\bf f}_{S}\left({x^{\prime}\over z},k^{\prime 2}\right)\nonumber\\
&-&{\bar \alpha_s}
\int_{x^{\prime}}^1 {dz\over z}
\int_{k^2}^{k^2/z}
{dk^{\prime 2}\over k^{\prime 2}}
\Biggl[\tilde {\frac{  {\bf F}_8 }{\omega^2}} \Biggr]
\Biggl(\frac{k^{\prime 2}}{k^2}z \Biggr)
\frac{ {\bf G}_0 }{2\pi^2}
{\bf f}_{S}\left({x^{\prime}\over z},k^{\prime 2}\right),
\label{dlress}
\eqnx }
where $\Biggl[\frac{\tilde {\bf F}_8}{\omega^2}\Biggr](z)$ denotes
the inverse Mellin transform of $\frac{{\bf F}_8}{\omega^2}$~:
\eq
\Biggl[\frac{\tilde {\bf F}_8}{\omega^2}\Biggr](z)=
\int_{\delta-i\infty}^{\delta+i\infty} {d\omega \over 2\pi i}
z^{-\omega}\frac{{\bf F}_8(\omega)}{\omega^2}.
\label{imellin}
\eqx	
with the integration contour located to the right of the singularities
of the function $\frac{{\bf F}_8(\omega)}{\omega^2}$.

To obtain $F_8$ we have to solve Eq.\ (\ref{f8}). The exact solution
presented in \cite{BARTNS,BARTS} may be expressed in terms of parabolic
cylinder functions~:
\eq
f_8^{\pm}=8\pi^2\bar\alpha_s {d \over d\omega} ln(e^{z^2/4} D_{p\pm}(z)),
\label{f8sol}
\eqx
where $f_8^{\pm}$ are two eigenvalues of matrix ${\bf F}_8(\omega)$ determined
in a basis of eigenvectors of ${\bf M}_8$, and~:
\eq
z={\omega \over \omega_0},\,\, \omega_0=\sqrt{N\bar \alpha_s}, \,\,
p_{\pm}={\lambda_8^{\pm} \over N},
\eqx
where $\lambda_8^{\pm}$ denote eigenvalues of matrix ${\bf M}_8$.

For a general solution (\ref{f8sol}) the inverse Mellin transform of
$\frac{{\bf F}_8(\omega)}{\omega^2}$
does not exist in the analytic form. However, it was checked in
\cite{BARTNS}
that approximate form of $F_8(\omega)$ determined in the large $N$ limit is a good
approximation for fixed $\alpha_s$. The large $N$
expansion proposed in \cite{BARTNS} reduces $F_8$ to the Born term.
In our case it implies~:
\eq
{\bf F}_8^{Born}(\omega)\approx 8\pi^2 \bar \alpha_s\frac{{\bf M}_8}{\omega}.
\eqx
The inverse Melin transform then reads~:
\eq
\Biggl[\frac{\tilde {\bf F}_8^{Born}}{\omega^2}\Biggr](z)=
4\pi^2 \bar \alpha_s{\bf M}_8 ln^2 (z).
\label{born}
\eqx
The different large $N$ approach which we propose here uses the approximate
form of ${\bf M}_0$, ${\bf M}_8$, ${\bf G}_0$ (\ref{m0}), (\ref{m8g0})~:
\eq
{\bf M}_0 \approx \left( \begin{array}{cc} N/2 & 0\\
                                           N   & 4N \\ \end{array} \right ),\,\,
{\bf M}_8 \approx \left( \begin{array}{cc} 0 & 0\\
                                           N & 2N \\ \end{array} \right ),\,\,
{\bf G}_0 \approx \left( \begin{array}{cc} N/2 & 0\\
                                           0   & N \\ \end{array} \right ),\,\,
\label{largen}
\eqx
where we have neglected all terms of the order less than $O(N)$.
Then the two components of $F_8$ simplify~:
\eqn
f_8^{+, large\,N}(\omega)&=&16\pi^2\frac{\omega\omega_0^2}{\omega^2-\omega_0^2}
,\nonumber\\
f_8^{-, large\,N}(\omega)&=&0,
\eqnx
and the inverse Mellin transform of $\frac{f_8^{+,large\,N}}{\omega^2}$ reads~:
\eq
\Biggl[\frac{\tilde f_8^{+,large\,N}}{\omega^2}\Biggr](z)=
8\pi^2 (z^{\omega_0}+z^{-\omega_0}-2 ).
\eqx

We have checked that both approaches give similar evolution of polarized
structure function $g_1$ and gluon distribution. Moreover, the branch
point singularities dominating the small $x$ behaviour of $g^{NS}_1$ and
$g^{S}_1$ behave also in  similar way. For the non-singlet case
the leading singularity may be determined from the Eq.\ (\ref{anomdres}) as~:
\eq
1-\frac{4 \bar\alpha_s }{\omega}
\Biggl(                 \frac{\dt P_{qq}(0)}{\omega} -
   \frac{( {\bf F}_8(\omega){\bf G}_0)_{qq} }{2\pi^2\omega^2} \Biggr)=0
\label{branchns}
\eqx
and for the ladder case it reads (cf. (\ref{lambdans}))~:
\eq
\lambda^{NS}=2\sqrt{\frac{N^2-1}{2N} \bar \alpha_s} \approx 0.39,
\eqx
where it was assumed that $\alpha_s=0.18$, $N=3$, $N_F=3$.
Including the non-ladder resummation gives~:
\eqn
\lambda^{NS,Born}     &\approx& 2\sqrt{\frac{N^2-1}{2N} \bar \alpha_s}
\Biggl(1+ \frac{1}{2 N^2}\Biggr) \approx 0.41,\nonumber\\
\lambda^{NS,large\,N} &\approx& 2\sqrt{\frac{N}{2} \bar \alpha_s} \approx
0.42.
\eqnx
The singlet dominating singularity fulfills the relation (cf. (\ref{anomdress}))~:
\eq
det\Biggl(
1-\frac{4 \bar\alpha_s }{\omega}
\Biggl(\frac{ {\bf M}_0 }{\omega}
      -\frac{ {\bf F}_8(\omega) {\bf G}_0 } {2\pi^2\omega^2} \Biggr) \Biggr)=0
\label{branchs}
\eqx
and for the ladder case it may be easily determined as~:
\eq
\lambda^{S}=2\sqrt{\lambda^+_0 \bar \alpha_s} \approx 1.13
\label{sin1}
\eqx
($\alpha_s=0.18$, $N=3$, $N_F=3$). The non-ladder resummation
changes the singularity point into~:
\eqn
\lambda^{S,Born}     &\approx& 1.01    \nonumber\\
\lambda^{S,large\,N} &\approx& 1.07.
\label{sin2}
\eqnx
One may notice that for both singlet and non-singlet case the influence
of the non-ladder resummation on the singularities is of the order 10\%.
Since the differences between Born and large N approaches are very small,
in the following discussion we restrict to the Born
approximation for ${\bf F}_8$.
\section{Unified evolution equation}

In order to make the quantitative predictions one has to constrain the
structure functions by the existing data at large and moderately small values
of $x$.  For such values of $x$ however the equations (\ref{dlres}) and
(\ref{dlress}) are inaccurate. In this region one should use the conventional
Altarelli-Parisi equations with complete splitting functions
$\Delta P_{ij}(z)$ and not restrict oneself to the effect generated only by
their $z\rightarrow 0$ part.
Following refs. \cite{BBJK,BZIAJA} we do therefore extend  equations
(\ref{dlres},\ref{dlress}) and
add to their right hand side the contributions coming from the
remaining parts of the splitting functions $\Delta P_{ij}(z)$.
We also allow the coupling $\alpha_s$ to run setting $k^2$ as the relevant
scale. In this way we obtain unified system of equations which contain
both the complete LO Altarelli-Parisi evolution and the double logarithmic
$ln^2(1/x)$ effects at low $x$.
The corresponding system of equations reads~:
{\small
\eqn
f_{k}(x^{\prime},k^2)=f^{(0)}_{k}(x^{\prime},k^2)
&+&
{\alpha_s(k^2)\over 2 \pi}{4\over 3}
\int_{x^{\prime}}^1 {dz\over z}
\int_{k_0^2}^{k^2/z}
{dk^{\prime 2}\over k^{\prime 2}}
f_{k}\left({x^{\prime}\over z},k^{\prime 2}\right)\label{unifns}\\
&&\hspace*{10ex}{\bf (\hspace*{3ex}Ladder\hspace*{3ex})}\nonumber\\
&+&{\alpha_s(k^2)\over 2\pi}\int_{k_0^2}^{k^2}{dk^{\prime 2}\over
k^{\prime 2}}{4\over 3}\int _{x^{\prime}}^1
{dz\over z} {(z+z^2)f_k({x^{\prime}\over z},k^{\prime 2})-
2zf_{k}(x^{\prime},k^{\prime 2})\over 1-z}\nonumber\\
&+&{\alpha_s(k^2)\over 2\pi}\int_{k_0^2}^{k^2}{dk^{\prime 2}\over
k^{\prime 2}}\left[2 +
{8\over 3} ln(1-x^{\prime})\right]f_{k}(x^{\prime},k^{\prime 2})\nonumber\\
&&\hspace*{10ex}{\bf(\hspace*{3ex}A-P\hspace*{4ex})}\nonumber\\
&-&{\alpha_s(k^2)\over 2\pi}
\int_{x^{\prime}}^1 {dz\over z}
\Biggl(
\Biggl[ \frac{\tilde  {\bf F}_8 }{\omega^2} \Biggr](z)
\frac{ {\bf G}_0 }{2\pi^2}
\Biggr)_{qq}
\int_{k_0^2}^{k^2}
{dk^{\prime 2}\over k^{\prime 2}}
f_{j}\left({x^{\prime}\over z},k^{\prime 2}\right)\nonumber\\
&-&{\alpha_s(k^2)\over 2\pi}
\int_{x^{\prime}}^1 {dz\over z}
\int_{k^2}^{k^2/z}
{dk^{\prime 2}\over k^{\prime 2}}
\Biggl(
\Biggl[\frac{\tilde   {\bf F}_8 }{\omega^2} \Biggr]
\Biggl(\frac{k^{\prime 2}}{k^2}z \Biggr)\frac{ {\bf G}_0 }{2\pi^2}
\Biggr)_{qq}
f_{j}\left({x^{\prime}\over z},k^{\prime 2}\right),\nonumber\\
&&\hspace*{10ex}{\bf(Non-ladder)}\nonumber
\eqnx
}
where $k=u_v, d_v, us$,
{\small
\eqn
f_{\Sigma}(x^{\prime},k^2)=f^{(0)}_{\Sigma}(x^{\prime},k^2)
&+&
{\alpha_s(k^2)\over 2 \pi}
\int_{x^{\prime}}^1 {dz\over z}
\int_{k_0^2}^{k^2/z}
{dk^{\prime 2}\over k^{\prime 2}}
{4\over 3}
f_{\Sigma}\left({x^{\prime}\over z},k^{\prime 2}\right)\nonumber\\
&-&{\alpha_s(k^2)\over 2 \pi}\int_{x^{\prime}}^1 {dz\over z}
\int_{k_0^2}^{k^2/z}
N_F{dk^{\prime 2}\over k^{\prime 2}}f_{g}
\left({x^{\prime}\over z},k^{\prime 2}\right)\nonumber\\
&&\hspace*{10ex}{\bf (\hspace*{3ex}Ladder\hspace*{3ex})}\nonumber\\
&+&{\alpha_s(k^2)\over 2 \pi}\int_{k_0^2}^{k^2}{dk^{\prime 2}\over
k^{\prime 2}}{4\over 3}\int _{x^{\prime}}^1
{dz\over z} {(z+z^2)f_{\Sigma}({x^{\prime}\over z},k^{\prime 2})-
2zf_{\Sigma}(x^{\prime},k^{\prime 2})\over 1-z}\nonumber\\
&+&{\alpha_s(k^2)\over 2 \pi}\int_{k_0^2}^{k^2}{dk^{\prime 2}\over
k^{\prime 2}}\left[ 2 +
{8\over 3} ln(1-x^{\prime})\right]
f_{\Sigma}(x^{\prime},k^{\prime 2}) \nonumber\\
&+&{\alpha_s(k^2)\over 2 \pi} \int_{k_0^2}^{k^2}{dk^{\prime 2}\over
k^{\prime 2}}\int _{x^{\prime}}^1
{dz\over z} 2z N_F f_g({x^{\prime}\over z},k^{\prime 2})\nonumber\\
&&\hspace*{10ex}{\bf(\hspace*{3ex}A-P\hspace*{4ex})}\nonumber\\
&-&{\alpha_s(k^2)\over 2\pi}
\int_{x^{\prime}}^1 {dz\over z}
\Biggl(
\Biggl[ \frac{\tilde  {\bf F}_8 }{\omega^2} \Biggr](z)
\frac{ {\bf G}_0 }{2\pi^2}
\Biggr)_{qq}
\int_{k_0^2}^{k^2}
{dk^{\prime 2}\over k^{\prime 2}}
f_{\Sigma}\left({x^{\prime}\over z},k^{\prime 2}\right)\label{unifsea}\\
&-&{\alpha_s(k^2)\over 2\pi}
\int_{x^{\prime}}^1 {dz\over z}
\int_{k^2}^{k^2/z}
{dk^{\prime 2}\over k^{\prime 2}}
\Biggl(
\Biggl[\frac{\tilde   {\bf F}_8 }{\omega^2} \Biggr]
\Biggl(\frac{k^{\prime 2}}{k^2}z \Biggr)\frac{ {\bf G}_0 }{2\pi^2}
\Biggr)_{qg}
f_{g}\left({x^{\prime}\over z},k^{\prime 2}\right),\nonumber\\
&&\hspace*{10ex}{\bf(Non-ladder)}
\nonumber
\eqnx
}
{\small
\eqn
f_{g}(x^{\prime},k^2)=f^{(0)}_{g}(x^{\prime},k^2) &+&
 {\alpha_s(k^2)\over 2 \pi}
\int_{x^{\prime}}^1 {dz\over z}
\int_{k_0^2}^{k^2/z}
{dk^{\prime 2}\over k^{\prime 2}}
{8\over 3}
f_{\Sigma}\left({x^{\prime}\over z},k^{\prime 2}\right)\nonumber\\
&+&{\alpha_s(k^2)\over 2 \pi}
\int_{x^{\prime}}^1 {dz\over z}
\int_{k_0^2}^{k^2/z}
{dk^{\prime 2}\over k^{\prime 2}}
12 f_{g}\left({x^{\prime}\over z},k^{\prime 2}\right)\nonumber\\
&&\hspace*{10ex}{\bf (\hspace*{3ex}Ladder\hspace*{3ex})}\nonumber\\
&+&{\alpha_s(k^2)\over 2 \pi}
\int_{k_0^2}^{k^2}
{dk^{\prime 2}\over k^{\prime 2}}\int_{x^{\prime}}^1 {dz\over z}
(-{4\over 3})zf_{\Sigma}
\left({x^{\prime}\over z},k^{\prime 2}\right)\nonumber\\
&+&{\alpha_s(k^2)\over 2 \pi}
\int_{k_0^2}^{k^2}
{dk^{\prime 2}\over k^{\prime 2}} \int_{x^{\prime}}^1 {dz\over z}
6z\left[{f_{g}
\left({x^{\prime}\over z},k^{\prime 2}\right)- f_{g}
(x^{\prime},k^{\prime 2})\over 1-z} -2f_{g}
\left({x^{\prime}\over z},k^{\prime 2}\right)\right]\nonumber\\
&+&{\alpha_s(k^2)\over 2 \pi}
\int_{k_0^2}^{k^2}
{dk^{\prime 2}\over k^{\prime 2}}\left[ {11\over 2} -{N_F\over 3}
 + 6 ln(1-x^{\prime})\right]f_{g}
(x^{\prime},k^{\prime 2})\nonumber\\
&&\hspace*{10ex}{\bf(\hspace*{3ex}A-P\hspace*{4ex})}\nonumber\\
&-&{\alpha_s(k^2)\over 2\pi}
\int_{x^{\prime}}^1 {dz\over z}
\Biggl(
\Biggl[ \frac{\tilde  {\bf F}_8 }{\omega^2} \Biggr](z)
\frac{ {\bf G}_0 }{2\pi^2}
\Biggr)_{gq}
\int_{k_0^2}^{k^2}
{dk^{\prime 2}\over k^{\prime 2}}
f_{\Sigma}\left({x^{\prime}\over z},k^{\prime 2}\right)\label{unifglue}\\
&-&{\alpha_s(k^2)\over 2\pi}
\int_{x^{\prime}}^1 {dz\over z}
\int_{k^2}^{k^2/z}
{dk^{\prime 2}\over k^{\prime 2}}
\Biggl(
\Biggl[\frac{\tilde   {\bf F}_8 }{\omega^2} \Biggr]
\Biggl(\frac{k^{\prime 2}}{k^2}z \Biggr)\frac{ {\bf G}_0 }{2\pi^2}
\Biggr)_{gg}
f_{g}\left({x^{\prime}\over z},k^{\prime 2}\right),\nonumber\\
&&\hspace*{10ex}{\bf(Non-ladder)}\nonumber
\eqnx
}
In equations (\ref{unifns}), (\ref{unifsea}), (\ref{unifglue})
we group separately terms corresponding to ladder 
diagram contributions to the double $ln^2(1/x)$ resummation, 
contributions from the non-singular parts of the Altarelli-Parisi 
splitting functions and finally contributions from the non-ladder 
bremmstrahlung diagrams.  We label those three contributions as 
"ladder", "AP" and "non-ladder" respectively.    

The inhomogeneous terms $f_i^{(0)}(x^{\prime},k^2)$ are expressed in terms of the input (integrated)
parton distributions and are the same as in the case of the LO Altarelli
Parisi evolution \cite{BBJK}:
\eqn
f_k^{(0)}(x^{\prime},k^2)&=&{\alpha_s(k^2)\over 2 \pi}{4\over 3}
\int _{x^{\prime}}^1
{dz\over z} {(1+z^2)\Delta p_k^{(0)}({x^{\prime}\over z})-
2z\Delta p_k^{(0)}(x^{\prime})\over 1-z}\nonumber\\
&&+{\alpha_s(k^2)\over 2 \pi}\left[2 +{8\over 3}
ln(1-x^{\prime})\right]\Delta p_k^{(0)}(x^{\prime})
\label{fns0}
\eqnx
($k=u_v,d_v,us$),

\eqn
f_{\Sigma}^{(0)}(x^{\prime},k^2) &=& {\alpha_s(k^2)\over 2 \pi}{4\over 3}
\int _{x^{\prime}}^1
{dz\over z} {(1+z^2)\Delta p_{\Sigma}^{(0)}({x^{\prime}\over z})-
2z\Delta p_{\Sigma}^{(0)}(x^{\prime})\over 1-z}\nonumber\\
&&+{\alpha_s(k^2)\over 2 \pi}(2 +{8\over 3} ln(1-x^{\prime}))
\Delta p_{\Sigma}^{(0)}(x^{\prime})\nonumber\\
&&+{\alpha_s(k^2)\over 2 \pi}N_F\int _{x^{\prime}}^1{dz\over z}
(1-2z)\Delta p_g^{(0)}({x^{\prime}\over z})\nonumber\\
f_g^{(0)}(x^{\prime},k^2) &=&{\alpha_s(k^2)\over 2 \pi}{4\over 3}
\int _{x^{\prime}}^1{dz\over z}(2-z)\Delta p_{\Sigma}^{(0)}({x^{\prime}\over
z})\nonumber\\
&&+{\alpha_s(k^2)\over 2 \pi}({11\over 2} -{N_F\over 3}
 + 6 ln(1-x^{\prime}))\Delta p_{g}^{(0)}(x^{\prime})\label{fsg0}\\
&&+{\alpha_s(k^2)\over 2 \pi}6
\int_{x^{\prime}}^1 {dz\over z}\left[
{\Delta p_{g}^{(0)}
({x^{\prime}\over z})- z\Delta p_{g}^{(0)}
(x^{\prime})\over 1-z} +(1-2z)\Delta p_{g}^{(0)}
({x^{\prime}\over z})\right].\nonumber
\eqnx
Equations (\ref{unifns}), (\ref{unifsea}), (\ref{unifglue}) together with
(\ref{fns0}), (\ref{fsg0}) and
(\ref{dpi}) reduce to the LO Altarelli-Parisi evolution equations with the
starting (integrated) distributions $\Delta p_i^{0}(x)$
after we set the upper
integration limit over $dk^{\prime 2}$ equal to $k^2$ in all terms in
equations (\ref{unifns}), (\ref{unifsea}), (\ref{unifglue}),
neglect the higher order terms in the kernels, 
and set $Q^2$ in place of $W^2$ as
the upper integration limit of the integral in Eq.\ (\ref{dpi}).

The presence of the running coupling changes the singularity
structure of the solution  turning the branch point
singularities into the infinite number of poles whose position
depends upon the magnitude of the cut-off $k_0^2$ \cite{JK}. 
Apparent branch point singularities are present if we adopt the
semiclassical approximation to the solution of equations
(\ref{unifns}), (\ref{unifsea}), (\ref{unifglue}) with the running 
coupling.  In this approximation we just
recover the RG structure with the running coupling i.e. 
\begin{equation} 
\left({k^2\over k_0^2}\right)^{\gamma_{NS}(\omega,\alpha_s)}
\rightarrow 
exp\left(\int_{k_0^2}^{k^2}{dk^{\prime 2}\over k^{\prime 2}}\gamma_{NS}
(\omega,\alpha_s(k^{\prime 2}))\right)
\label{rgstr}
\end{equation}  
and similarly for the singlet part.

Introducing the Altarelli-Parisi kernel into the double logaritmic evolution
equations (\ref{dlres}),(\ref{dlress}) changes the anomalous dimensions $\gamma$
(\ref{anomdres}), (\ref{anomdress}). They take then the form~:
\eqn
\gamma_{NS}(\omega,\alpha_s ) &=&
{\omega+H_{qq}(\omega) - \sqrt{(\omega-H_{qq}(\omega))^2 - 4 P_{qq}(\omega)\omega}\over 2}
\label{anomdap}\nonumber\\
{\tdm\gamma}_{S}(\omega,\alpha_s) &=& {\omega+{\bf H}(\omega) -
\sqrt{ (\omega-{\bf H}(\omega) )^2 - 4 {\bf P}(\omega)\omega}\over 2},
\label{anomdsap}
\eqnx
where $H(\omega)$ denotes the moment representation of non-singular part
of the Altarelli-Parisi kernel,
and $P(\omega)$ denotes the moment of double logaritmic kernel i.\ e.\
for ladder case $P(\omega)={\bar \alpha_s {\bf M}_0 \over \omega}$.
We investigated the influence of the Altarelli-Parisi kernel on the behaviour
of leading sigularities for the non-singlet and gluonic sector (see Tab. \
1), assuming for simplicity~:
\eqn
{\bf H}(\omega) &\approx& {\bf H}(\omega=0)\nonumber\\
{\bf P}(\omega) &=&\frac{{\bf P}_1}{\omega}-\frac{{\bf P}_2}{\omega^3},
\eqnx
where ${\bf P}_1(\omega)=\bar \alpha_s {\bf M}_0$ and
${\bf P}_2(\omega)=4 \bar \alpha_s^2 {\bf M}_8{\bf G}_0$.

In Table 1 we summarize  numerical results concerning the
position of leading  branch points for the non-singlet and 
singlet case obtained in  different approximations of the
kernel i.e.  for the pure double logarithmic approximation
generated by ladder diagrams alone (first column), for the
complete double logarithmic approximation (second column) and
finally for the complete double logaritmic approximation
supplemented by the non-singular part(s) od the
Altarelli-Parisi splitting functions (third and fourth column). 
One may notice
that  the position of the branch point for the non-singlet
part is to a very good accuracy determined by the
double logarithmic approximation
generated by ladder diagrams alone, and the non-ladder
contributions and the non-singular parts of the AP kernel do not
play important role.  The latter terms are however important in
the singlet case, and in particular  the non-singular part
of the Altarelli-Parisi splitting function significantly reduce
the magnitude of the position of the branch point singularity.  
This means that in the singlet case the corrections to the
double logarithmic approximation may be expected to be very
important.  This effect will be quantified in the next Section 
where we present results of the numerical solution(s) of
equations (\ref{unifns}), (\ref{unifsea}), (\ref{unifglue}).

\begin{table}[hbpt]
\noindent
Table 1.\ Leading singularities for the non-singlet and gluonic part of
$g_1$ determined for $\alpha_s=0.18$, $N=3$, $n_f=3$.
\begin{center}
\begin{tabular}{|l|c|c|c|c|c|c|c|}
\hline \hline
                    &$ladder$                  &$+non-ladder$&&$ladder+(A-P)$&$+non-ladder$\\
                    &                          &             &&              &$+(A-P)$     \\
                    &$H\equiv 0$, $P_2\equiv0$&$H\equiv 0$  &&$P_2\equiv0$&\\
\hline
$\lambda^{NS}$      &$0.39$                    &$0.41$       &&$0.41$&$0.43$\\
\hline
$\lambda^{S} $      &$1.17$                    &$1.08$       &&$0.96$&$0.78$\\
\hline \hline
\end{tabular}
\end{center}
\end{table}
\subsection{Numerical results}

We solved equations (\ref{unifns}), (\ref{unifsea}), (\ref{unifglue})
assuming the following simple parametrisation of the
input distributions:

\begin{equation}
\Delta p_i^{(0)}(x)=N_i (1-x)^{\eta_i}
\label{dpi0}
\end{equation}
with $\eta_{u_v}=\eta_{d_v}=3,$  $\eta_{\bar u} = \eta_{\bar s} = 7 $
and $\eta_g=5$.  The normalisation constants $N_i$ were determined
 by imposing the Bjorken sum-rule for $\Delta u_v^{(0)}-\Delta d_v^{(0)}$
 and requiring that the first moments of
all other distributions are the same as those determined from  the recent
QCD analysis \cite{STRATMAN}. All distributions $\Delta p_i^{(0)}(x)$
behave as $x^0$ in the limit $x\rightarrow 0$ that corresponds to the implicit
assumption that the Regge poles which
should control the small $x$ behaviour of $g_1^{(0)}$ have their intercept equal
to $0$.
\newline
\begin{figure}[htb]
   \vspace*{-1cm}
    \centerline{
     \psfig{figure=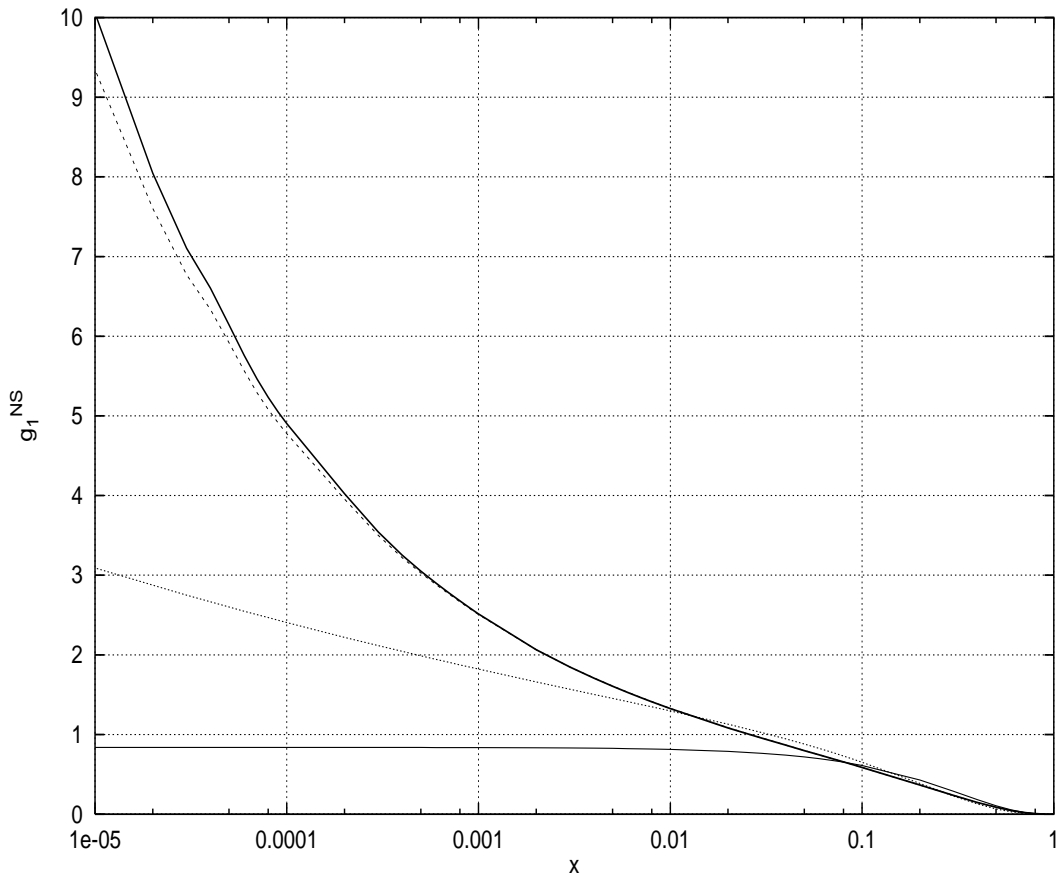,height=12cm,width=14cm}
               }
    \vspace*{-0.5cm}
     \caption{Non-singlet part of the proton spin structure function $g_1(x,Q^2)$
as a function of $x$ for $Q^2=$10 GeV$^2$. Solid line
corresponds to the calculations which contain the full
$ln^2(1/x)$ resummation with both bremsstrahlung corrections and 
Altarelli-Parisi kernel included, 
dashed line represents the ladder $ln^2(1/x)$ resummation with
Altarelli-Parisi kernel included, a dotted line shows the pure
Altarelli-Parisi evolution, and a thin solid one describes the input
non-perturbative part  $g_1^{(NS, 0)}$.} 
\label{fig.2}
\end{figure}
\newline
\begin{figure}[htb]
   \vspace*{-1cm}
    \centerline{
     \psfig{figure=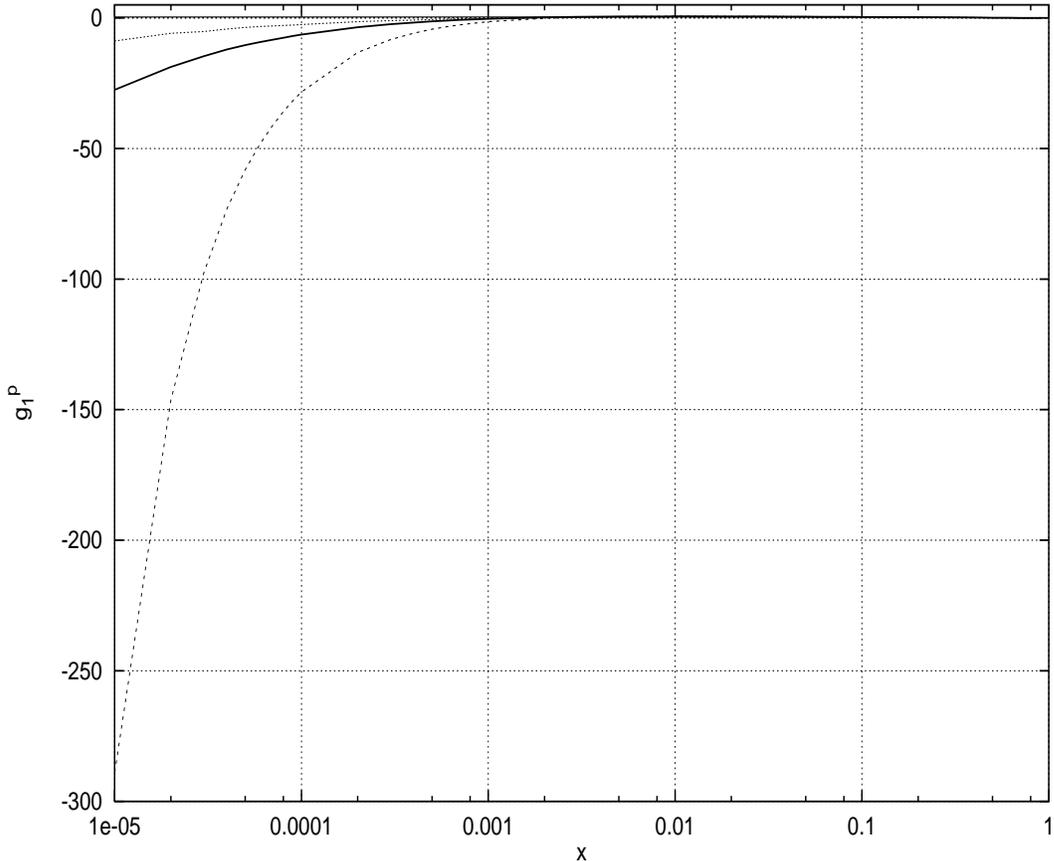,height=12cm,width=14cm}
               }
    \vspace*{-0.5cm}
\caption{ The structure function $g_1^p(x,Q^2)$ for $Q^2=10 GeV^2$ plotted as
the function of $x$.  
Solid line
corresponds to the  calculations which contain the full
$ln^2(1/x)$ resummation with both bremsstrahlung corrections and Altarelli-Parisi
kernel included, 
dashed line represents the ladder $ln^2(1/x)$ resummation with
Altarelli-Parisi kernel included, a dotted line shows the pure
Altarelli-Parisi evolution, and a thin solid one describes the input
non-perturbative part  $g_1^{(p, 0)}$.}
\label{fig.3}
\end{figure}
\newline
\begin{figure}[htb]
   \vspace*{-1cm}
    \centerline{
     \psfig{figure=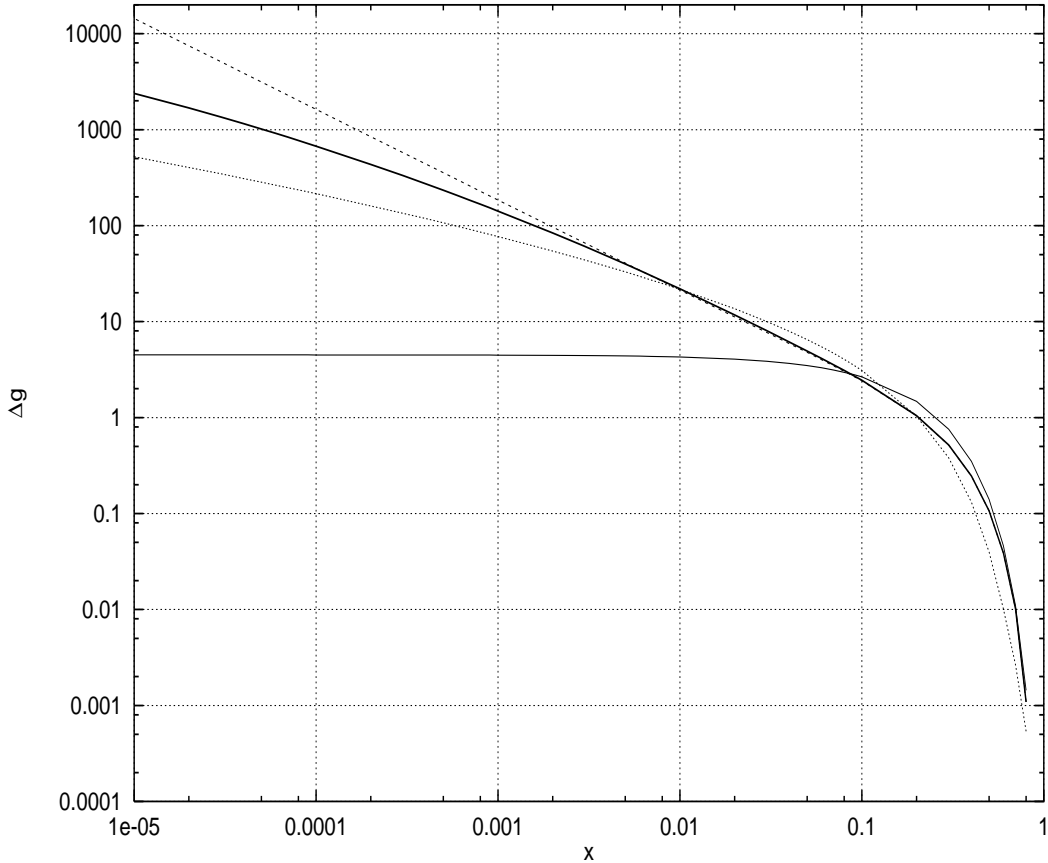,height=12cm,width=14cm}
               }
    \vspace*{-0.5cm}
     \caption{The spin dependent gluon distribution $\Delta G(x,Q^2)$
for $Q^2=10 GeV^2$ plotted as the function
of $x$.  
Solid line
corresponds to the  calculations which contain the full
$ln^2(1/x)$ resummation with both bremsstrahlung corrections and Altarelli-Parisi
kernel included, 
dashed line represents the ladder $ln^2(1/x)$ resummation with
Altarelli-Parisi kernel included, a dotted line shows the pure
Altarelli-Parisi evolution, and a thin solid one describes the input
non-perturbative part  $\Delta G^{(0)}$. }
\end{figure}
It was checked that the parametrisation (\ref{dpi0}) combined with
equations
(\ref{dpi}), (\ref{gp1}),(\ref{unifns}), (\ref{unifsea}), (\ref{unifglue}) gives
reasonable description of the recent SMC data on $g_1^{NS}(x,Q^2)$ and on
$g_1^p(x,Q^2)$ \cite{SMC}.
In Fig.\ 5 we present the nonsinglet part of $g_1(x,Q^2)$
for $Q^2= 10 GeV^2$ in the small $x$
region \cite{BBJK}. We show predictions based on   equations
(\ref{dpi},\ref{unifns}) and confront them with the results
obtained from the solution of the LO Altarelli-Parisi evolution equations 
with the input distributions at $Q_0^2=1 GeV^2$ given by equations 
(\ref{dpi0}). 
We also show results based on equations similar to (\ref{unifns}) in
which we have removed  the bremsstrahlung contributions to the
kernel.  We may see from this figure that the double logarithmic
contributions are very important at low $x$, and that they are
reasonably well described by the contribution of ladder
diagrams. 
We also plot the nonperturbative part of the non-singlet distribution
$g_1^{NS(0)}(x)=g_A/6(1-x)^3$, where $g_A$ is the axial vector coupling.

In Fig.\ 6 we again confront predictions for $g_1^p(x,Q^2)$ at $Q^2=10 GeV^2$ 
based on equations (\ref{dpi}, \ref{unifsea}, \ref{unifglue})
with those based on the LO Altarelli-Parisi evolution equations, 
and with the results based on  equations similar to (\ref{unifsea}, \ref{unifglue}) 
in which we have removed  the bremsstrahlung contributions to the
kernel.
In the region of very low values of $x$ the dominant contribution comes from 
the singlet component of $g_1^p$.   
We see from this figure that the contribution  of the bremsstrahlung term  is 
very important and significantly slows down the increase of $g_1^p$ with 
decreasing $x$. 
The structure function $g_1^p(x,Q^2)$ which contains effects
of the double $ln^2(1/x)$ resummation begins to differ from that calculated within
the LO Altarelli Parisi equations already for $x \sim 10^{-3}$.

In Fig.\ 7 we show the spin dependent gluon distribution which
contains effects of the double $ln^2(1/x)$ resummation
and confront it with that which was obtained
from the LO Altarelli-Parisi equations.  
It can be seen that the ladder resummation exhibits characteristic
$x^{-\lambda_S}$ behaviour with $\lambda_S \sim 1$.  Similar behaviour is also
exhibited by the structure function $g_1^p(x,Q^2)$ itself.
The contribution  of the bremsstrahlung term  is 
very important and significantly slows down the increase of
$\Delta G$ with decreasing $x$.

\section{Conclusions}

To sum up we have presented theoretical expectations for the low $x$ behaviour of
the spin dependent structure function $g_1(x,Q^2)$ which follows from the
resummation of the double $ln^2(1/x)$ terms.
We have also presented results of the analysis of the "unified" equations
which contain the LO Altarelli Parisi evolution and the double $ln^2(1/x)$ effects
at low $x$.  
We based our calculation on a formalism of the unintegrated
spin dependent distributions which satisfied the corresponding
integral equations.  This formalism made it possible to make an 
insight into physical origin of the double $ln^2(1/x)$
resummation.  The structure of the corresponding equations is similar to
the conventional LO Altarelli - Parisi evolution equations with
extended kernels to account for the bremsstrahlung contributions.
Very important difference however is the absence of
transverse momenta ordering along the chain  (let us remind that
the LO Altarelli-Parisi evolution  corresponds to ladder diagrams 
with ordered transverse momenta).  This should have important
implications for the structure of the final state in polarized
deep inelastic scattering at low $x$,  similarily to the case of
unpolarized DIS \cite{FSTATE}.          
\section*{Acknowledgments}

This research has been supported in part by the Polish Committee for Scientific
Research grants 2 P03B 184 10, 2 P03B 89 13 and 2P03B 04214.

\end{document}